\documentclass[a4paper]{article}
\usepackage{fullpage}
\usepackage{amsmath,amssymb,theorem}
\usepackage{hyperref}
\usepackage[nosort]{cite}

\let\frac\undefined

\allowdisplaybreaks

\numberwithin{equation}{section}

\def\Maketitle{{\def\newpage{}\maketitle}}

\def\eq#1{\begin{equation}#1\end{equation}}

\def\Align#1{\begin{align}#1\end{align}}

\def\Aligned#1{\begin{aligned}#1\end{aligned}}

\def\Gathered#1{\begin{gathered}#1\end{gathered}}
\def\Multline#1{\begin{multline}#1\end{multline}}

\def\Cases#1{\begin{cases}#1\end{cases}}
\def\?{\notag}
\def\d{\partial}
\def\bd{{\bar\partial}}
\def\bL{{\bar L}}

\def\Res{\mathop{\rm Res}}

\def\const{\mathop{\rm const}\nolimits}
\def\cA{{\cal A}}
\def\bcA{{\cal\bar A}}

\def\cD{{\cal D}}
\def\bcD{{\cal\bar D}}

\def\cO{{\cal O}}

\def\sh{\mathop{\rm sh}\nolimits}
\def\ch{\mathop{\rm ch}\nolimits}
\def\th{\mathop{\rm th}\nolimits}

\def\lcolon{\mathopen{\,:\,}}
\def\rcolon{\mathclose{\,:\,}}

\def\Z{{\mathbb{Z}}}

\def\cP{{\cal P}}
\def\cQ{{\cal Q}}

\def\e{{\rm e}}
\def\i{{\rm i}}

\def\bz{{\bar z}}

\def\bc{{\bar c}}
\def\bh{{\bar h}}

\makeatletter

\def\section{\@startsection{section}{1}{\z@}%
                                   {-3.5ex \@plus -1ex \@minus -.2ex}%
                                   {2.3ex \@plus.2ex}%
                                   {\normalfont\normalsize\bfseries}}
\def\subsection{\@startsection{subsection}{2}{\z@}%
                                     {-3.25ex\@plus -1ex \@minus -.2ex}%
                                     {1.5ex \@plus .2ex}%
                                     {\normalfont\normalsize\bfseries\itshape}}
\def\@seccntformat#1{\csname the#1\endcsname.~~}
\long\def\@makecaption#1#2{%
  \vskip\abovecaptionskip
  \sbox\@tempboxa{\small#1. #2}%
  \ifdim \wd\@tempboxa >0.9\hsize
  {\leftskip=0.05\hsize\rightskip=0.05\hsize\relax\small
    #1. #2\par}
  \else
    \global \@minipagefalse
    \hb@xt@\hsize{\hfil\box\@tempboxa\hfil}%
  \fi
  \vskip\belowcaptionskip}
\def\Appendix{\appendix
  \def\@seccntformat##1{Appendix~\csname the##1\endcsname.~~}}

\let\over\@@over
\let\atop\@@atop
\let\above\@@above
\let\overwithdelims\@@overwithdelims
\let\atopwithdelims\@@atopwithdelims
\let\abovewithdelims\@@abovewithdelims

\makeatother

\begin{document}

\title{On form factors and Macdonald polynomials}
\author{Michael Lashkevich and Yaroslav Pugai\\
\parbox[t]{.9\textwidth}{
\normalsize\it\begin{center}
Landau Institute for Theoretical Physics,\\
142432 Chernogolovka, Russia%
%
%
\\
and\\
Moscow Institute of Physics and Technology\\
141707 Dolgoprudny of Moscow Region, Russia
\end{center}
}}
\date{}

\Maketitle

\begin{abstract}
We are developing the algebraic construction for form factors of local operators in the sinh-Gordon theory proposed in~\cite{Feigin:2008hs}. We show that the operators corresponding to the null vectors in this construction are given by the degenerate Macdonald polynomials with rectangular partitions and the parameters $t=-q$ on the unit circle. We obtain an integral representation for the null vectors and discuss its simple applications.
\end{abstract}

\section{Introduction}
We study the structure of the space of local fields in integrable two-dimensional massive models by developing the algebraic approach proposed in\cite{Feigin:2008hs}. Our basic example, the massive sinh-Gordon model, is defined by the action
\eq{
\cA=\int d^2x\,\left({1\over 16\pi}(\d_\mu\varphi)^2-2\lambda\ch b\varphi\right).
\label{shG-action}
}
The space of local operators of the model consists of the exponential operators $\e^{\alpha\varphi}$ and fields of the form
\eq{
\d^{n_1}\varphi\cdots\d^{n_k}\varphi\,\bd^{m_1}\varphi\cdots
\bd^{m_k}\varphi\,\e^{\alpha\varphi}\,,
\label{descend-def}
}
where $\d={1\over2}(\d_1-\d_0)$, $\bd={1\over2}(\d_1+\d_0)$ are derivatives corresponding to the light-cone variables $z=x^1-x^0$, $\bz=x^1+x^0$. The pair of the numbers $(l,l')$, where $l=\sum n_i$, $l'=\sum m_i$, is called the level of the operator. The number $s=l-l'$ is its (Lorentz) spin. An operator with $l'=0$ will be called a right chiral descendant operator, while that with $l=0$ will be called a left one.

When the coupling constant $\lambda$  is zero and the action describes the free massless boson field, the operators (\ref{descend-def}) are well-defined and form a basis in the space of local operators. The problem we address is the structure of the space of local operators in the interactive case. To elucidate the problems we encounter recall that the sinh-Gordon model can be considered as a perturbation of the Liouville field theory: $\cA=\cA_{\rm Liouv.}+\cA_{\rm pert.}$, where
\eq{
\Aligned{
\cA_{\rm Liouv.}
&=\int d^2x\,\left({1\over 16\pi}(\d_\mu\varphi)^2-\lambda\e^{b\varphi}\right),
\\
\cA_{\rm pert.}
&=-\lambda\int d^2x\,\e^{-b\varphi}.
}\label{Liouville}
}
The Liouville theory is a model of the conformal field theory\cite{Belavin:1984vu}. The Laurent components $L_k$ of the energy-momentum tensor form the Virasoro algebra with the central charge $c=1+6Q^2$, $Q=b+b^{-1}$. We assume $L_k$ to act on any local operator $\cO(z,\bz)$ according to the rule
\eq{
L_k\cO(z,\bz)=\oint{dw\over2\pi\i}(w-z)^{n+1}T(w)\cO(z,\bz),
\label{Lk-def}
}
where
\eq{
T(z)=-{1\over4}(\d\varphi)^2+{Q\over2}\d^2\varphi.
\label{Tz-def}
}
Here the integration is taken over a small circle around the point $x=(z,\bz)$ in the Euclidean plane.

The exponential fields are primary operators with respect to the Virasoro algebra. The conformal dimension of the field $\e^{\alpha\varphi}$ is $\Delta_\alpha=\alpha(Q-\alpha)$. In fact, the fields $\e^{\alpha\varphi}$ and $\e^{(Q-\alpha)\varphi}$, which have the same dimensions, are proportional in the Liouville theory\cite{Zamolodchikov:1995aa}: $\e^{\alpha\varphi}=R(\alpha)\e^{(Q-\alpha)\varphi}$. The reflection coefficient $R(\alpha)$ is known exactly. This fact is related to the exponentially growing potential term in the Liouville action.

Equations (\ref{Lk-def}), (\ref{Tz-def}) define the action of the Virasoro algebra on the Fock spaces with the bases~(\ref{descend-def}). At generic values of $\alpha$ these spaces are irreducible representations of the Virasoro algebra, but for particular values of $\alpha$ their structure as Virasoro representations is more complicated. For example, it is not difficult to check that
\eq{
\cD_{12}\e^{\alpha\varphi}|_{\alpha\to-b/2}=0,
\label{l2-zero}
}
where
$$
\cD_{12}=L_{-1}^2+b^2L_{-2}.
$$
The only level~2 vector that belongs to the irreducible module generated from the field $\e^{-(b/2)\varphi}$ is $\d^2\e^{-{b\over2}\varphi}=L_{-1}^2\e^{-{b\over2}\varphi}$. Take any linearly independent level~2 operator, e.g.\ $(\d\varphi)^2\e^{-{b\over2}\varphi}$. Such operator corresponds to a so called cosingular vector in the representation of the Virasoro algebra. The cosingular vector is defined modulo the vector corresponding to $L_{-1}^2\e^{-(b/2)\varphi}$ and generates the whole Virasoro module. The cosingular vector operator can be obtained as well by means of a limiting procedure. Namely, the operator
$$
\left.{1\over\alpha+b/2}\cD_{12}\e^{\alpha\varphi}\right|_{\alpha\to-b/2}
=-b\left((\d\varphi)^2-Q\d^2\varphi\right)\e^{-(b/2)\varphi}
$$
is its another representative.

Now consider the Fock space over the `reflected' field~$\e^{(Q+b/2)\varphi}$. Its structure is different. The field $\cD_{12}\e^{\alpha\varphi}$ does not vanish as $\alpha\to Q+b/2$. On the contrary, it gives an operator
\eq{
\cD_{12}\e^{(Q+b/2)\varphi}
=Q(Q+b)\left((\d\varphi)^2+b\d^2\varphi\right),
\label{12-singular}
}
which is apparently nonzero. This operator corresponds to a singular (or null) vector and possesses some peculiar features. The correlation functions of this singular vector operator vanish, if all other operators are the exponential operators or their Virasoro descendants. In particular,
$$
\left\langle\e^{-(b/2)\varphi(x_1)}\cD_{12}\e^{(Q+b/2)\varphi(x_2)}\right\rangle=0.
$$
But if a correlation function contains a cosingular vector operator or a Virasoro descendant of a cosingular vector operator, it can be nonzero. For example,
$$
\left\langle(\d\varphi)^2\e^{-(b/2)\varphi(x_1)}\cD_{12}\e^{(Q+b/2)\varphi(x_2)}\right\rangle
={\const\over(z_1-z_2)^{2\Delta_{12}+4}(\bz_1-\bz_2)^{2\Delta_{12}}}
$$
with a nonzero proportionality constant.

The described structure is general for the operators $V_{mn}(z,\bz)=\e^{\alpha_{mn}\varphi}$ with $\alpha_{mn}={1-m\over2}b^{-1}+{1-n\over2}b$. There are operators $\cD_{mn}$ of the level $m\cdot n$, constructed from the Virasoro operators $L_{-k}$ ($k>0$), such that $\cD_{mn}V_{mn}=0$. The codimension of the level $mn$ subspace of the Virasoro module generated from~$V_{mn}$ is just one, so that there is only one (modulo this subspace) cosingular vector. The `reflected' descendant operator $\cD_{mn}V_{-m,-n}\ne0$ is a singular vector one. The singular vectors are known to be represented in terms of the Jack symmetric polynomials corresponding to rectangular ($n\times m$) partitions\cite{Awata:1994xd}. The symmetric power (Newton) polynomials must be substituted by the boson oscillator modes. We shall see that there are analogous vectors in the form factor theory given by the Macdonald polynomials with rectangular partitions and special values of parameters, $q=-t=\e^{-\i\pi r}$, with a similar substitution for the symmetric power polynomials.

The existence of the singular and cosingular vector operators leads to two important facts. The first fact is the possibility to reduce the Liouville theory by excluding these operators from the theory. More precisely, the reduced theory only contains the operators that belong to the irreducible Virasoro modules generated over exponential operators. The resulted theory is minimal in the sense that it contains just one primary field of each conformal dimension. The classes of primary and of exponential operators coincide in such reduced Liouville theory. Surely, from the physical point of view this reduction is negligible on the background of the continuous spectrum of the Liouville field theory. However, it becomes essential, if the theory is restricted to the degenerate operators, as it takes place in the Liouville gravity or in the minimal $c<1$ conformal models.

The second fact is more subtle and is related to the existence of the second Virasoro algebra, related to the second light-cone variable~$\bz$. We shall denote its generators as~$\bL_k$ and the operators that annihilate $V_{mn}$ as $\bcD_{mn}$. Define the operators
\eq{
\Phi_{\alpha|mn}(z,\bz)={\cD_{mn}\bcD_{mn}\e^{\alpha\varphi(z,\bz)}\over(\alpha-\alpha_{mn})^2}.
\label{Phi-def}
}
It is easy to show that the operator $\Phi_{\alpha|mn}$ is expressed as a linear combinations of the basic operators~(\ref{descend-def}) with the coefficients finite as $\alpha\to\alpha_{mn}$. Nevertheless, in the Liouville field theory it possesses a pole at this point due to the resonance phenomenon\cite{Zamolodchikov:1990bk}. The singular part of the field is proportional to $V_{m,-n}(z,\bz)$, while the regular one gives the renormalized version of this operator:
\eq{
\Phi_{\alpha|mn}(z,\bz)={2\Lambda^{4nb(\alpha-\alpha_{mn})}B_{mn}\e^{(\alpha-nb)\varphi(z,\bz)}\over\alpha-\alpha_{mn}}
+\Phi^{\rm ren}_{mn}(\Lambda|z,\bz)+O(\alpha-\alpha_{mn}).
\label{Phi-resonance}
}
Here $\Lambda$ is an arbitrary parameter of the dimension of mass, which affects the definition of the renormalized operator. It reflects the uncertainty due to the logarithmic divergences in the perturbation theory. The singular part was found exactly (including the coefficient~$B_{mn}$) by Al.~Zamolodchikov in his prominent work on the higher equations of motion in the Liouville theory\cite{Zamolodchikov:2003yb}.

It is important to notice that the resonance identity (\ref{Phi-resonance}) preserve the same form in the sinh-Gordon theory unless the integers $m$ and $n$ are both odd\cite{Lashkevich:2011ne}. In the case of odd $m,n$ the r.h.s.\ of the resonance identity is modified by a contribution of additional operators. Here we show that the corresponding resonance identities for even $mn$ exist in the bootstrap form factor theory. A more complete and mathematically rigorous consideration will be published separately.

Return to the sinh-Gordon theory. Recall that it can be considered as a perturbation of the Liouville theory in two ways: we may consider the term with $\e^{b\varphi}$ as a part of the Liouville action and that with $\e^{-b\varphi}$ as a perturbation, and vice versa. The second way differs from (\ref{Liouville}) by the substitution $\varphi\to-\varphi$. Hence, there are two kinds of identifications in the sinh-Gordon theory:
\eq{
\Aligned{
\e^{\alpha\varphi}
&=R(\alpha)\e^{(Q-\alpha)\varphi},
\\
\e^{\alpha\varphi}
&=R(-\alpha)\e^{-(Q+\alpha)\varphi}.
}\label{ShG-reflections}
}
Repeating these identities we obtain an infinite set of identities:
\eq{
G_{\beta}\e^{\alpha\varphi}=G_\alpha\e^{\beta\varphi},\quad
\text{if ${\beta-\alpha\over Q}\in2\Z$ or ${\alpha+\beta\over Q}\in2\Z+1$.}
\label{VV-equiv}
}
Here $G_\alpha=\langle\e^{\alpha\varphi}\rangle$ is the vacuum expectation value of the exponential operator~\cite{Lukyanov:1996jj}.

Surely, the Virasoro algebra is not an algebra of conserved currents in the sinh-Gordon theory. Nevertheless, there are some traces of the singular and cosingular vectors there. The multiple identifications~(\ref{VV-equiv}) look for the degenerate values of $\alpha$ as follows:
\eq{
G_{m'n'}V_{mn}=G_{mn}V_{m'n'},
\text{if $m'-m=n'-n\in4\Z$ or $m+m'=n+n'\in4\Z$.}
\label{VVmn-equiv}
}
Naturally, up to the overall normalization the form factors of these operators coincide. We shall see below that the construction of the form factors contains some objects that correspond to the singular and cosingular vectors of all equivalent operators~$V_{mn}$.

Off the conformal point there is no infinite-dimensional Virasoro like symmetry in the coordinate space. Still the theory is integrable and it has an infinite number of commutative integrals of motion. This leads to factorized scattering and admits a bootstrap approach.

Namely, we treat the theory (\ref{shG-action}) as a scattering theory that contains the only sort of particles with a factorized $S$ matrix. The two-particle $S$ matrix reads
\eq{
S(\theta)={\th{1\over2}(\theta+\i\pi(r-1))\over\th{1\over 2}(\theta-\i\pi (r-1))}\,,
}
where $\theta$ is the rapidity difference and the parameter $r$ is defined by the relation $b^2=(1-r)/r$. In this bootstrap theory every local operator $\cO(x)$ is defined by a set of form factors, i.e.\ the matrix elements $\langle{\rm vac}|\cO(x)|\theta_1,\ldots,\theta_N\rangle$ between the vacuum state and the eigenstate containing $N$ particles with the rapidities $\theta_1,\ldots,\theta_N$, being considered as analytic functions of the rapidities. All other matrix elements are expressed in terms of these ones by means of the crossing symmetry. The set of the form factors for any operator is a solution to a system of bootstrap equations (the so called form factor axioms)\cite{Karowski:1978vz,Smirnov:1992vz}. Any solution corresponds to some local operator. Nevertheless, it is yet unknown how to identify particular solutions to these equations with operators defined in terms of the Lagrangian fields. In particular, in the case of the sinh-Gordon theory the identification is known for the exponential operators $\e^{\alpha\varphi}$\cite{Koubek:1993ke,Lukyanov:1997bp} and several other operators (see e.g.~\cite{Smirnov:1992vz}). There is no general identification for the descendant operators~(\ref{descend-def}). Here we shall show that, nevertheless, it is possible to find some traces of the Lagrangian field theory, in particular, the resonance poles and resonance identities, in the form factor theory. It provides some reference points for a future identification procedure.

Note that the sine-Gordon model can be defined by the same action (\ref{shG-action}) but with an imaginary parameter $b$. The spectrum of the sine-Gordon model consists of kinks and breathers. The particle of the sinh-Gordon theory transforms into the first breather of the sine-Gordon model for imaginary values of~$b$. All discussed below is applicable to the sine-Gordon model being restricted to the breather sector.

The problem of the one-to-one correspondence between the operator spaces in conformal field theory and massive integrable models in the framework of the bootstrap form factor approach has a long history. The correspondence was first explicitly demonstrated in the case of the Ising model in~\cite{Cardy:1990pc}. Some evidence for the one-to-one correspondence for a more general case was found in~\cite{Koubek:1994di}. Rigorous proofs for some rather general cases were found in~\cite{Babelon:1996xq,Babelon:1996sk,Jimbo:2003ge,Jimbo:2003ne}. In the particular case of the Lee--Yang model a simple proof was proposed in~\cite{Delfino:2005wi,Delfino:2007bt}. Some promising steps towards finding the explicit correspondence has been made in~\cite{Jimbo:2010jv,Jimbo:2011bc}. The importance of resonances for understanding this problem was noted in~\cite{Zamolodchikov:1990bk,Belavin:2003pu} for the Lee--Yang model and in~\cite{Fateev:2006js,Fateev:2009kp} for the $Z_N$ Ising model.

In this paper we only give some most important results without computational details and provide their elementary explanations. A more detailed and extended treatment of the subject will be provided in the forthcoming paper.

The paper is organized as follows. In section~\ref{sect:freefield} we recall the basic results of~\cite{Feigin:2008hs}. In section~\ref{sect:nullvectors} we explain what can be called null vectors in the form factor theory and formulate the main statement concerning their relation to the Macdonald polynomials. In section~\ref{sect:screening} we introduce the screening operators and give the integral representation for the degenerate Macdonald polynomials with rectangular partitions. In section~\ref{sect:resonance} we show how the resonance phenomenon manifests in the bootstrap form factors.

\section{Free field construction for generic \texorpdfstring{$\alpha$}{alpha}}
\label{sect:freefield}

The form factors in the sinh-Gordon model were studied by many authors\cite{Zamolodchikov:1990bk,Fring:1992pt,Koubek:1993ke,Lukyanov:1997bp}. It was shown that they as functions of the rapidities $\theta_j$ factorize as follows
\eq{
\langle{\rm vac}|\cO(0)|\theta_1,\ldots,\theta_N\rangle
=\rho^NJ_{\cO}(\e^{\theta_1},\ldots,\e^{\theta_N})\prod_{i<j}R(\theta_i-\theta_j).
\label{Fhh'-def}
}
Here $\cO(x)$ is any local field. The constant
\eq{
\rho=\left(2\cos{\pi r\over2}\right)^{-1/2}\exp\int^{\pi(1-r)}_0{dt\over2\pi}\,{t\over\sin t}
\label{rho-def}
}
and the function
\eq{
R(\theta)=\exp\left(
4\int^\infty_0{dt\over t}\,
{\sh{\pi t\over2}\sh{\pi (r-1)t\over2}\sh{\pi rt\over2}
\over\sh^2\pi t}\ch(\pi-\i\theta)t
\right),
\label{R-def}
}
often called the minimal form factor, constitute a factor common for all operators in the theory. The operator-specific factor $J_{\cO}(x_1,\ldots,x_N)$ is a rational symmetric function of the variables $x_i=\e^{\theta_i}$ with the only poles at $x_i=-x_j$. For a definite spin $s$ operator this factor is also homogeneous of the homogeneity power~$s$. Below we concentrate on the study of this rational factor.

The idea of the work \cite{Feigin:2008hs} was to construct the functions $J_\cO(x_1,\ldots x_N)$ algebraically. Consider two sets of the operators  $d^+_k$ and $d^-_k$ ($k\ne0$) that commute within each set,
\eq{
\Aligned{{}
[d^+_k,d^+_l]
&=0\,,
\\
[d^-_k,d^-_l]
&=0\,,
}\label{dOsc-def}
}
and have the following non-trivial relations between the two sets:
\eq{
[d^+_k,d^-_l]=kA_k\delta_{k+l,0}\,.
\label{dOscNontr-def}
}
Here the additional factor
\eq{
A_k=-4\i^{-k}\sin{\pi k(r-1)\over2}\sin{\pi kr\over2}.
\label{A-def}
}
is chosen to simplify the expressions below. Define the highest weight vectors $|a\rangle$ and $\langle a|$. By definition, the oscillators  $d^+_k$ and $d^-_k$ with  positive $k$ annihilate the ket highest weight vector, while those with negative $k$ annihilate the bra highest weight vector. The highest weight vectors are eigenvectors of the zero mode $\cP$:
\Align{
&\cP|a\rangle=a|a\rangle,
\qquad
d^\pm_k|a\rangle=0\,,
\label{rvac-def}
\\
&\langle a|\cP=\langle a|a,
\qquad
\langle a|d^\pm_{-k}=0
\quad(k>0).
\label{lvac-def}
}
With the Heisenberg algebra operators (\ref{dOscNontr-def}) let us construct the current similar to that of the deformed Virasoro algebra:
\eq{
t(z)=e^{\i\pi\cP}\lambda_-(z)+e^{-\i\pi\cP} \lambda_+(z),
\label{BosDVA}}
where the operators $\lambda_+$ and $\lambda_-$ have an exponential form in the oscillators:
\eq{
\lambda_\pm(x)=\exp\sum_{k\ne0}{d^\pm_kz^{-k}\over k}\,.
\label{lambdapm-def}
}
The construction (\ref{dOsc-def})---(\ref{lambdapm-def}) was  adjusted in such a way that it give a simple free field representation for the form factors of the exponential operators. Let%
\footnote{The definition of $a$ here differs by the sign from that of~\cite{Feigin:2008hs}. Due to the reflection relations it is physically equivalent, but the correspondence between the descendant fields and the elements of the algebra~$\cA$ changes.}
$$
V_a(x)=\e^{\alpha\varphi(x)},
\qquad
a={1\over2}-{\alpha\over Q}.
$$
The rational part of the form factors of the operator $G_a^{-1}V_a$ is given by
\eq{
J_a(x_1,\ldots,x_N)=\langle a|t(x_1)\cdots t(x_N)|a\rangle\,.
\label{PrimPol}
}
This matrix element is easily computed by means of the simple rule
\eq{
\Gathered{
\lambda_\pm(z)\lambda_\pm(w)=\lcolon\lambda_\pm(z)\lambda_\pm(w)\rcolon,
\\
\lambda_+(z)\lambda_-(w)=\lambda_-(w)\lambda_+(z)=f(w/z)\lcolon\lambda_+(z)\lambda_-(w)\rcolon,
}\label{lambda-norm}
}
where $f(z)={(z+q)(z-q^{-1})/(z^2-1)}$.

The approach makes it possible not only to compute form factors for the exponential fields but also for descendants. In the case of generic values of the parameters $r$ and $a$ the naive guess would be to find them by computing arbitrary matrix elements of the product $t(x_1)\cdots t(x_N)$. However, an explicit check showed the resulting expressions not to satisfy the kinematic pole condition in general. It was found that the Fock space vectors that lead to form factors of local fields are given in terms of the following combinations of oscillators
\eq{
\Aligned{
&c_{-k}={d^-_k-d^+_k\over A_k},\\
&\bc_{-k}={d^-_{-k}-d^+_{-k}\over A_k}
\quad(k>0).
\label{pibpi-def}
}
}
where the denominator $A_k$ is defined in~(\ref{A-def}). The elements $c_{-k}$, $\bc_{-k}$ commute except for the even elements of different sorts:
\eq{
[c_{-2k},\bc_{-2k}]=-4kA_{2k}^{-1}.
\label{cbc-commut}
}
Besides, evidently,
\eq{
c_{-k}|a\rangle=0,
\qquad
\langle a|\bc_{-k}=0.
\label{cbc-vac}
}
The elements $c_{-k}$ form a commutative algebra $\cA$ with the natural grading $\cA=\bigoplus^\infty_{l=0}\cA_l$: the element $c_{-k}$ has the degree~$k$. The dimension of the level $l$ subspace $\cA_l$ coincides with the number of level $l$ chiral descendants. The vectors~$\langle a|h$, $h\in\cA$, form a free right module of the algebra~$\cA$. It can be proved that each matrix element $\langle a|h\,t(x_1)\cdots t(x_N)|a\rangle$ with $h\in\cA_l$ gives the rational part of the form factors of some level $(l,0)$ right descendant operator. Moreover, for generic values of $r$ and $a$ every right chiral descendant operator can be represented in such form. Let $\bcA$ be the commutative algebra generated by the elements $\bc_{-k}$. If $h\in\cA$, the element $\bh\in\bcA$ is obtained by the substitution $c_{-k}\to\bc_{-k}$. The vectors $\bh|a\rangle$ form a free left module of~$\cA$. The matrix elements $\langle a|t(x_1)\cdots t(x_N)\,\bh|a\rangle$ with $\bh\in\bcA_l$ produce form factors of a level $(0,l)$ left chiral descendant operator.

A generic matrix element
\eq{
J^{h\bh'}_a(x_1,\ldots,x_N)=\langle a|h\,t(x_1)\ldots t(x_N)\,\bh'|a\rangle
\label{Jhha-def}
}
corresponds to some descendant operator~$V^{h\bh'}_a(x)$ of spin $l-l'$, if $h\in\cA_l$, $h'\in\cA_{l'}$. It can be shown that if $h\in\cA_l$, $h'\in\cA_{l'}$ the resulting operator is a linear combination of descendants of levels not exceeding $l$ in the right sector and $l'$ in the left one. The conjecture that its level is just $(l,l')$ has not been well substantiated, though not yet disproved as well.

The functions $J^{hh'}_a$ are computed by means of pushing $h$ to the right and $\bar h'$ to the left with the use of the commutation relations
\eq{
\Aligned{{}
[c_{-k},\lambda_\pm(z)]
&=(\mp)^{k+1}z^k\lambda_\pm(z),
\\
[\lambda_\pm(z),\bc_{-k}]
&=(\pm)^{k+1}z^{-k}\lambda_\pm(z).
}\label{clambda-commut}
}
This rule together with (\ref{lambda-norm}), (\ref{cbc-commut}), (\ref{cbc-vac}) uniquely defines the matrix elements~(\ref{Jhha-def}).

By definition the function $J^{h\bh'}_a$ is quasiperiodic with respect to the variable~$a$:
\eq{
J^{h\bh'}_{a+1}(x_1,\ldots,x_N)=(-)^NJ^{h\bh'}_a(x_1,\ldots,x_N).
\label{quasiperiodicity}
}
This quasiperiodicity reflects a part of the reflection equations. For the exponential operators ($h=h'=1$) another symmetry property appears: $J_{-a}=J_a$. Together with (\ref{quasiperiodicity}) it forms all the reflection properties and the property that the substitution $\alpha\to-\alpha$ ($a\to1-a$) is equivalent to the substitution $\varphi\to-\varphi$. For generic $h,h'$ the reflection $a\to-a$ demands also application of some map (reflection map) to the elements $h,h'$\cite{Feigin:2008hs}. We may say that the symmetries that follow from both perturbation theories (about the free massless boson and about the Liouville theory) are realized in the algebraic construction in a nontrivial way, while a combination of two such symmetries is only realized trivially. This fact stresses the nonperturbative nature of form factors.

For generic values of $r$ and $\alpha$ the definition (\ref{Jhha-def}) provides the existence of a (yet unknown explicitly) one-to-one correspondence between elements of $\cA\otimes\bcA$ and descendant operators~(\ref{descend-def}). For special values of $\alpha$ it is not so. We shall see that, at least, for the values $\alpha=\pm\alpha_{mn}$ there are some singular or cosingular vectors that break this correspondence.

\section{Null vectors and Macdonald polynomials}
\label{sect:nullvectors}

We have already said that in the conformal field theory the Fock space is not irreducible as a representation of the Virasoro algebra. Having in mind further application of the construction to the restricted sine-Gordon model we would like to study possible reductions in the space of local fields.

Let us focus our attention on one (right) chirality and consider the form factors of the level~2 descendant operators. There are two elements $c_{-2}$ and $c_{-1}^2$, which produce such operators. The explicit computation for the $V^{c_{-2}}_a$ operator gives the following expressions up to three particles:
\eq{
\Aligned{
\langle a|{c}_{-2}|a\rangle
&=0\,,
\\
\langle a|{c}_{-2}\>t(x_1)|a\rangle
&=2\i \sigma_1^2\sin\pi a\,,
\\
\langle a|{c}_{-2}\>t(x_1)t(x_2)|a\rangle
&=2\i \bigl((\sigma_1^2-2\sigma_2)\sin2\pi a-2\sigma_2\sin\pi r\bigr)\,,
\\
\langle a|{c}_{-2}\>t(x_1)t(x_2)t(x_3)|a\rangle
&={8\i \sigma_1^2\sigma_3\sin^2\pi r\sin\pi a\over\prod_{i<j}(x_i+x_j)}
\\
&\quad+4\i\cos\pi a\,((\sigma_1^2-2\sigma_2)\sin2\pi a-2\sigma_2\sin\pi r)\,.
}\label{c2ff}
}
Here $\sigma_r=\sigma_r(x_1,\ldots,x_N)$ denotes the $r$th elementary symmetric polynomial. As for the operator $V^{c_{-1}^2}_a$, its form factors are proportional to the form factor of the exponential field $V_a$ with the extra factor $\sigma_1^2=\left(\sum x_j\right)^2$. The corresponding matrix elements are
\eq{
\Aligned{
\langle a|{c}_{-1}^2|a\rangle&=0\,,
\\
\langle a|{c}_{-1}^2\>t(x_1)|a\rangle&=2\sigma_1^2\cos\pi a\,,
\\
\langle a|{c}_{-1}^2\>t(x_1)t(x_2)|a\rangle&=4 \sigma_1^2 \cos^2 \pi a\,,
\\
\langle a|{c}_{-1}^2\>t(x_1)t(x_2)t(x_3)|a\rangle
&=4\sigma_1^2\cos\pi a\,\biggl({\sigma_1\sigma_2-\sigma_3\cos2\pi r\over\prod_{i<j}(x_i+x_j)}
+\cos2\pi a\biggr)\,.
}\label{c11ff}
}
For generic exponential fields $V_a$ the expressions for the two level~2 operators $V^{c_{-2}}_a$ and $V^{c_{-1}^2}_a$ are linearly independent. However, it is easy to see that for $a=a_{12}\equiv1-{r\over 2}$ the following identity holds
$$
\langle a_{12}|{c}_{-2}\>t(x_1)\cdots t(x_N)|a_{12}\rangle=
-\i\tan{\pi r\over 2}\>\langle a_{12}|{c}_{-1}^2\>t(x_1)\cdots t(x_N)|a_{12}\rangle\,.
$$
for $N=0,\ldots,3$. It also can be checked for an arbitrary nonnegative integer~$N$. Since the identity between polynomials holds for all matrix elements, this is a relation between the corresponding local fields. In other words, for the second level descendants of the exponential field $V_{a_{12}}$ a nonvanishing null (or singular) vector
$$
N_{12}\sim c_{-2}+\i\tan{\pi r\over2}\>c_{-1}^2\,
$$
appears, such that
$$
\langle a_{12}|N_{12}t(x_1)\cdots t(x_N)|a_{12}\rangle=0
$$
for arbitrary $N$, $x_i$, and, hence, $V^{N_{12}}_{a_{12}}=0$.

A level 2 nonzero null vector appears as well for $a=a_{21}\equiv1-{1-r\over2}$:
$$
N_{21}\sim c_{-2}+\i\cot{\pi r\over 2}\>c_{-1}^2\,,
$$
so that
$$
\langle a_{21}|N_{21}t(x_1)\cdots t(x_N)|a_{21}\rangle=0\,.
$$
This observation can be continued further and one can check that similar identities appear for all $a=a_{mn}\equiv{1\over2}(mr+n(1-r))$. These values correspond to the values $\alpha=\alpha_{mn}$ in the sense that $V_{a_{mn}}=V_{mn}=\e^{\alpha_{mn}\varphi}$.

It can be shown that the singular vectors are expressed in terms of the Macdonald symmetric functions. Let $P_{\lambda;q,t}(\{p_k\}^\infty_{k=1})$ be the Macdonald symmetric function (see also Appendix) for a partition~$\lambda$ and parameters $q,t$ as a function of the symmetric power (Newton) polynomials $p_k=\sum x_i^k$. Let $n\times m$ be the rectangular partition $\{m\}^n_{i=1}$. For the values
\begin{equation}
q=-t=\exp(-\i\pi r)
\label{qtspecific}
\end{equation}
define the algebraic element
\eq{
N_{mn}=C_{mn}P_{n\times m;q,t}(\{c_{-k}(1-q^{-k})\})\in\cA.
\label{Pmn-Macdonald}
}
We reserved an arbitrary nonzero coefficient $C_{mn}$ for future convenience. This element gives a singular vector in the following sense:
\eq{
\langle a_{mn}|N_{mn}\>t(x_1)\cdots t(x_N)|a_{mn}\rangle =0\,,
\label{maineq}
}
for all sets $\{x_1,\ldots,x_N\}$. The corresponding form factor vanishes as well. The quasiperiodicity property (\ref{quasiperiodicity}) leads to the identification of some Fock spaces, so that the Fock space over $\langle a_{mn}|$ contains infinitely many singular vectors in the form factor sense.%
\footnote{Such abundance of null vectors is the specificity of sinh-Gordon model related to the reflection equations. We expect that in the sine-Gordon model the null vectors $\langle a_{mn}|N_{mn}$ only survive. It seems very plausible that these singular vectors reflect the existence of the cosingular vectors in the Liouville theory.}
Namely, we have
\eq{
\langle a_{mn}|N_{m-n+s,s}\>t(x_1)\cdots t(x_n)|a_{mn}\rangle =0\,,
\qquad
s>0,n-m;\ s-n\in2\Z.
\label{svect-gen}
}
The result for the left chiral sector is obtained by the substitution $a_{mn}\to a_{-m,-n}$:
\eq{
\langle a_{-m,-n}|\>t(x_1)\cdots t(x_n)\bar N_{m-n+s,s}|a_{-m,-n}\rangle =0\,,
\qquad
s>0,n-m;\ s-n\in2\Z.
\label{barsvect-gen}
}
We see that the construction does not respect the $P$-invariance ($z\leftrightarrow-\bz$) of the theory. Instead, it respects the combined symmetry $z\leftrightarrow-\bz$, $\alpha\to Q-\alpha$.

Due to the $\varphi\to-\varphi$ symmetry of the sinh-Gordon theory, we may expect the same nullification over the exponential operators~$\e^{-\alpha_{mn}\varphi}$. Again, the algebraic construction for form factors turns out not to respect this symmetry. Instead of singular vectors for the corresponding values $a=1-a_{mn}$ we have a kind of cosingular vectors. We shall describe this structure in detail elsewhere. Here we give examples of such cosingular vectors later in section~\ref{sect:resonance}, since the cosingular vectors in the left sector play an important role in the resonance identities.

\section{Screening operators}
\label{sect:screening}

The integral representation for the Macdonald polynomials with rectangular partitions for generic values of $q,t$ is known to be related to the integral representation of null vectors of the deformed Virasoro algebra in terms of the screening operators\cite{Shiraishi:1995rp}. This representation does not have the limit $t\to-q$, regular on the unit circle, due to the divergence of the $q$-gamma functions and the theta functions it involves. Here we present a construction of the screening operators and an integral representation in the case $t=-q$.

The basic object of the integral representation is the screening current%
\footnote{This expression appears as a $x\to\i$ limit of the deformed screening current defined in\cite{Lukyanov:1994re} with $q=x^{-2r}$, $t= x^{-2(r-1)}$.}
\eq{
\Aligned{
S(z)
&=\e^{\i(r-1)\cQ}\exp\left(\sum_{k>0}B_k\bc_{-k}z^k\right)
\exp\left(\sum_{k>0}B_kc_{-k}z^{-k}\right),
\\
B_k
&={q^{k/2}-(-)^kq^{-k/2}\over k}\,.
}\label{S-def}
}
Here we introduced the zero mode $\cQ$ canonically conjugate to the operator $\cP$: $[\cP,\cQ]=-\i$.  The screening current satisfies the following commutation relation
$$
\cP S(z)=S(z)(\cP+r-1).
$$
Recall that in our construction the zero mode $\cP$ only enters the definition of the particle creation operator~(\ref{BosDVA}). For this reason we will identify the vectors
\eq{
\Aligned{
\e^{i (r-1)\cQ}|a_{mn}\rangle
&=|a_{m,n-2}\rangle\,,\cr
\langle a_{m,n-2}|\e^{\i(r-1)\cQ}
&=\langle a_{mn}|\,,
}\label{ashift}
}
and, thus, the screening current shifts the eigenvalues of~$\cP$.

By using the commutation relations for the operators $d^\pm_k$ we find the exponential fields to satisfy the following operator expansion
\Align{
S(z)S(w)
=\left(1-{w^2\over z^2}\right)\lcolon S(z)S(w)\rcolon.
\label{Ssigma-prod}
}
It means that the modes of the screening current essentially anticommute. Now we introduce the screening operators composed of a `fermion' single screening operator, $\Sigma$, and a `boson' double screening operator $W$, quadratic in the current\footnote{It is interesting that these operators looks very similar to those defined in \cite{Babelon:1996sk} for sine-Gordon model at the reflectionless points.}:
\Align{
\Sigma_{mn}
&=\oint{dz\over2\pi\i}\,z^{m-n}S(z),
\label{Sigma-def}
\\
W_{mn}
&=\oint{dz_1\over2\pi\i}\oint{dz_2\over2\pi\i}\,
z_1^{m-n+2}z_2^{m-n}S(z_1)S(z_2)F^n(z_2/z_1)\,,
\label{W-def}
}
where the factor $F^n(z)$ is a formal series
\eq{
F^n(z)=\sum^\infty_{k=1}F^n_kz^k=\sum^\infty_{k=1}(-)^{k}{1-(-)^nq^{k}\over 1+(-)^nq^{k}}z^k.
\label{Fn-def}
}
The operators $\Sigma$ and $W$ make it possible to define the operator
\eq{
Q^{(s)}_{mn}=\Cases{W_{m,n-2s+4}\cdots W_{m,n-4}W_{m,n},&\text{if $s\in2\Z$;}\\
\Sigma_{m,n-2s+2}W_{m,n-2s+6}\cdots  W_{m,n-4}W_{m,n},&\text{if $s\in2\Z+1$.}}
\label{Q-def}
}
The form of the operator depends drastically on the evenness of the integer~$s$. An important property of the operator is its commutation with the current~$t(z)$, if $s-n$ is even:
\eq{
[Q^{(s)}_{mn},t(z)]=0
\text{ on the Fock space over $|a_{mn}\rangle$, if $s-n\in2\Z$.}
\label{Qt-commut}
}
Besides
\eq{
Q^{(s)}_{mn}|a_{mn}\rangle=0,
\text{ if $s>n-m$.}
\label{Q1-zero}
}
Hence, we have the identity
\eq{
\langle a_{m,n-2s}|Q^{(s)}_{mn}t(x_1)\cdots t(x_N)|a_{mn}\rangle=0.
\label{Qtt-zero}
}
Define the element $N_{mn}$ in terms of the operator $Q$ according to
\eq{
\langle a_{mn}|N_{mn}=\langle a_{m,-n}|Q^{(n)}_{mn}.
\label{Q-gen:def}
}
Then eq.~(\ref{svect-gen}) is the immediate consequence of~(\ref{Qtt-zero}). We used the arbitrariness of $C_{mn}$ to fix the normalization of the null vector. While performing the computation of the r.h.s.\  one can find that only a finite number of terms of the series (\ref{Fn-def}) enters the answer and the expressions are well defined for generic values of the parameter~$r$. The result of taking integrals will be a sum of the terms $c_{-k_1}\cdots c_{-k_s}$ with $k_1+...+k_s=mn$. The identity (\ref{maineq}) (as well as (\ref{svect-gen})) immediately follows from (\ref{Qt-commut}) and~(\ref{Q1-zero}). To check the connection of the singular vectors (\ref{Q-gen:def}) with the degenerate Macdonald polynomials it is necessary to check that the properties described in Appendix are satisfied.

This provides the promised integral representation for the degenerate Macdonald polynomials:
\Multline{
P_{n\times m;q,-q}=C_{mn}^{-1}
\Res_{w_1=0}\cdots\Res_{w_{n-1}=0}
\prod^{n-1}_{i=1}w_i^{-(n-i)(m+i)-1}\prod_{1\le i\le j<n}\left(1-\prod^j_{t=i}w_t^2\right)
\prod^{\lfloor n/2\rfloor}_{i=1}F^n(w_{2i-1})\times
\\*
\times\Res_{z=0}z^{-mn-1}\exp\sum^\infty_{k=1}{q^{k/2}-(-)^kq^{-k/2}\over k(q^{k/2}-q^{-k/2})}
q^{k/2}\left(1+\sum^{n-1}_{i=1}\prod^i_{j=1}w_j^k\right)z^kp_k.
\label{Pss'-int}
}
The last residue in this expression just extracts a finite number of terms of the power~$mn$. The integral representation is valid, in particular, on the unit circle $|q|=1$ off roots of unity. The constants $C_{mn}$ must be chosen so that it provide the correct coefficient e.g.\ at $p_1^{mn}$. Up to now we have no explicit general expression for these constants.

\section{Resonance identities}
\label{sect:resonance}

Note that the construction of null vectors for the left chiral sector is essentially the same as for the right one, but a singular vector appears for $a=a_{-m,-n}$, $m,n>0$, (see (\ref{barsvect-gen})), while for $a=a_{mn}$ a cosingular vector appears. It leads to a new phenomenon, when we study descendant fields that contain both chiralities or, in our language, when both the $c_{-k}$ and  $\bc_{-k}$ operators meet.

Consider as an example the vector $\langle a_{m1}|N_{m1}$ with even $m>0$. By definition (\ref{Q-gen:def}) $\langle a_{m1}|N_{m1}=\langle a_{m,-1}|\Sigma_{m1}$. The operator $\Sigma$ has a nontrivial commutation relation with the left chirality oscillator. For even $m$ the commutator $[c_{-m},\bc_{-m}]\ne0$ and we have the following exact relation
$$
\Sigma_{m1}\bc_{-m}|a_{m1}\rangle=\kappa_m|a_{m,-1}\rangle,
\qquad
m\in2\Z\,,
$$
where $\kappa_m=-{2/(q^{m/2}-q^{-m/2})}$. One may say that $c_{-m}|a_{m1}\rangle$ represents a cosingular vector, which is dual to the singular vector $\langle a_{m1}|N_{m1}$. Indeed, consider the matrix element:
\Multline{
\langle a_{m1}|N_{m1}t(x_1)\cdots t(x_k)\bc_{-m}|a_{m1}\rangle
\\
=\langle a_{m,-1}|\Sigma_{m,1}t(x_1)\cdots t(x_k)\bc_{-m}|a_{m1}\rangle
=\langle a_{m,-1}|t(x_1)\cdots t(x_k)|\Sigma_{m,1}\bc_{-m}|a_{m1}\rangle
\\
=\kappa_m\>\langle a_{m,-1}|t(x_1)\cdots t(x_k)|a_{m,-1}\rangle.
\label{Pm1-id}
}
The matrix element in the first line corresponds to the field $G_{m1}^{-1}V^{N_{m1}\bar{c}_{m}}_{m1}$, while that in the last line to the exponential field~$G_{m,-1}^{-1}V_{m,-1}$. Hence, these two operators coincide. Note that the operator $V^{N_{m1}\bc_{-m}}_a$ is a level $(m,m)$ descendant. It means that the identity (\ref{Pm1-id}) provides an identity between a descendant field and an exponential field with different values of~$a$.

More generally, define the element $M_{mn}\in\cA$ by
\eq{
\bar M_{mn}|a_{mn}\rangle=(\kappa^{(m+n)}_m)^{-1}Q^{(m)}_{m,2m+n}|a_{m,2m+n}\rangle,
}
where
$$
\kappa^{(2k)}_m=k!\prod^k_{i=1}F^m_{2i-1},
\qquad
\kappa^{(2k+1)}_m=(-)^kk!\prod^k_{i=1}F^{m+1}_{2i}.
$$
Notice that this element does not coincide with $N_{2n-m,n}$ and is not represented in terms of the Macdonald polynomials unless $m-n$ is even. It can be seen that
$$
Q^{(n)}_{mn}\bar M_{mn}|a_{mn}\rangle=(\kappa^{(m+n)}_m)^{-1}Q^{(m+n)}_{m,2m+n}|a_{m,2m+n}\rangle=|a_{m,-n}\rangle,
$$
and, hence,
\eq{
\langle a_{mn}|N_{mn}t(x_1)\cdots t(x_k)\bar M_{mn}|a_{mn}\rangle
=\langle a_{m,-n}|t(x_1)\cdots t(x_k)|a_{m,-n}\rangle,
\quad\text{if $mn\in2\Z$.}
\label{NM-id}
}
This means the operator identity $G_{mn}^{-1}V^{N_{mn}\bar M_{mn}}_{mn}(x)=G_{m,-n}^{-1}V_{m,-n}(x)$. Comparing this relation with (\ref{Phi-resonance}), which, as we have already mentioned, remains valid in the sinh-Gordon model for even values of $mn$, we obtain
$$
\Phi_{\alpha|mn}(x)={2B_{mn}\over\alpha-\alpha_{mn}}{G_{m,-n}\over G_{mn}}V^{N_{mn}\bar M_{mn}}_{1/2-\alpha/Q}(x)+O(1).
$$
We see that this is a rather weak identity. It only fixes the poles of the coefficients in the expansion of the field $\Phi_{\alpha|mn}(x)$ in the level $(mn,mn)$ descendants obtained by $c_{-k}$, $\bc_{-k}$ from the field $V_a(x)$. Nevertheless, up to now it is nearly the only available information concerning the descendant operators.

\section{Conclusion}

Here we give a brief report of some results obtained within the algebraic approach to form factors of descendant fields. In particular, for the sinh-Gordon model there are vectors analogous to the null vectors in the Liouville theory among the descendants of the exponential fields $\exp(\alpha_{mn}\varphi)$. We prove that such vectors can be expressed in terms of the Macdonald polynomials with the parameters $t=-q$ on the unite circle. We construct an integral representation for these polynomials and show that the null vectors might be important in studying the space of local fields. As an application of the construction we consider a set of operator resonances in the sinh-Gordon theory. We show that the corresponding resonance identities can be obtained in the framework of the algebraic approach.

It would be interesting to clarify the relation of the proposed construction to the fermion structure in the sine-Gordon model discovered in~\cite{Jimbo:2010jv,Jimbo:2011bc}. The advantage of the approach by Jimbo et al.\ is that it is based on the lattice model, where all operators are well-defined, and the renormalization only enters the construction during the scaling limit. This makes it possible to relate the operators in the conformal field theory and in the massive model. A half of the fermion modes in their construction seems to be related to the modes of our screening current.

A more detailed study of the level $(2,2)$ descendants is also necessary. It is the first nontrivial example, where both chiralities are present. In particular, the results should be compared with~\cite{Delfino:2006te}. We delay it for a separate paper.

\section*{Acknowledgments}

We are grateful to A.~Belavin, M.~Bershtein, B.~Feigin, F.~Smirnov for many useful discussions and for the encouragement. M.~L.\ is thankful to LPTHE, Universit\'e Pierre et Marie Curie and, especially, to F.~Smirnov for the hospitality during his visit in October 2012 and LIA Physique Th\'eorique et Mati\`ere Condens\'ee for the financial support of this visit. The study was supported, in part, by the Ministry of Education and Science of Russian Federation under the contracts \#8410 and~\#8528 and by Russian Foundation for Basic Research under the grant \#13-01-90614.

\appendix
\section{Macdonald polynomials for generic \texorpdfstring{$q,t$}{q,t}}

The Macdonald polynomials and the Macdonald operator are defined in~\cite{Macdonald:SFHP}. Consider the space of symmetric polynomials from variables $\{x_1, \ldots, x_N\}$ for some natural $N$. The ring of symmetric polynomials can be generated by the symmetric power functions
$$
p_k=\sum_{i=1}^Nx_i^k\,.
$$
It is convenient to use a basis in this ring indexed by partitions. Let $\lambda=\{\lambda_i\}^\infty_{i=1}$ be a partition, i.e.\ a sequence of nonnegative integers $\lambda_1\ge\lambda_2\ge\cdots\ge0$ such that $|\lambda|=\sum^\infty_{i=1}\lambda_i$ is finite. Then $p_\lambda=\prod^\infty_{i=1}p_{\lambda_i}$.

It is possible to introduce an inner product in this ring. For the polynomials $p_k$ we define
\eq{
\langle p_k,p_l\rangle=k{1-q^k\over1-t^k}\delta_{kl}\,,
\label{pp-scal}
}
which depends on a couple of parameters $q$ and $t$. For the general polynomials $p_\lambda$ we assume
\eq{
\langle p_\lambda,p_\mu\rangle
=\delta_{\lambda\mu}S_\lambda\prod^\infty_{i=1}\langle p_{\lambda_i},p_{\lambda_i}\rangle\,,
\label{pp-part-scal}
}
where
$$
S_\lambda=\prod^\infty_{k=1}\#\{i\in\Z_{>0}|\lambda_i=k\}!
$$
is the number of the admissible permutations of the partition~$\lambda$. The factor $S_\lambda$ in the inner product imposes the Wick rule on the symmetric power polynomials, making possible to define the boson operators $\hat p_k$,~$\d\over\d p_k$ on the space of symmetric polynomials.

The Macdonald polynomials are known to be eigenfunctions of the following difference operator (the Macdonald operator)
$$
H=\sum_{i=1}^N\prod_{j\neq i}^N{tx_i-x_j\over x_i-x_j}T_{q,i}\,,
$$
where the operator $T_{q,i}$ acts on any function of the variables $\{x_i\}$ by dilation of the $i$th variable:
$$
T_{q,i} f(x_1,\ldots,x_i,\ldots,x_N)=f(x_1,\cdots,qx_i,\cdots,x_N).
$$

Suppose the length of the partition $\lambda$ does not exceed the number of variables~$N$. In other words, $\lambda_{N+1}=0$. The Macdonald polynomial $P_{\lambda}=P_\lambda(x_1,\ldots, x_N)$ is the degree $|\lambda|$ symmetric polynomial that satisfies the condition
$$
HP_{\lambda}=\sum_{i=1}^Nt^{N-i}q^{\lambda_i}P_\lambda.
$$

It is convenient to consider the Macdonald polynomials with a large enough number $N$ of variables as a function of the symmetric power functions~$p_k$: $P_\lambda=P_{\lambda;q,t}(p_1,p_2,\ldots)$. These functions are orthogonal with respect to the scalar product~(\ref{pp-scal})---(\ref{pp-part-scal}). The Macdonald polynomials are usually normalized as follows:
$$
\langle P_\lambda,P_\mu\rangle=\delta_{\lambda\mu}
\prod_{s\in\lambda}{1-q^{a(s)+1}t^{l(s)}\over1-q^{a(s)}t^{l(s)+1}},
$$
where the product is taken over cells of the Young tableau corresponding to the partition $\lambda$, and $a(s)$ and $l(s)$ are the `arm' and `leg' lengths of the cell~$s$.

The simplest examples of the Macdonald polynomials are $P_\varnothing=1$, $P_{1}=p_1$ and the two polynomials used in section~\ref{sect:nullvectors}:
$$
\Aligned{
P_{11}
&={1\over2}(-p_2+p_1^2).
\\
P_{2}
&={(1-q)(1+t)p_2+(1+q)(1-t)p_1^2\over2(1-qt)},
}
$$
The last two functions give the elements $N_{12}$, $N_{21}$ introduced in Sect.~\ref{sect:nullvectors} after the specialization $t=-q=\e^{-\i\pi r}$ and the substitution $p_k\to c_{-k}(1-q^{-k})$. For generic $N_{mn}$ the proof of (\ref{Pmn-Macdonald}) is based on the boson representation (i.e.\ the representation in terms of the operators $\hat p_k$, $\d\over\d p_k$) of the Macdonald operator obtained in~\cite{Awata:1994xd}.

\raggedright

\end{document}